\documentclass[twocolumn,amssymb,amsmath,nobibnotes, aps, pra]{revtex4-1}
\usepackage{graphicx}
\usepackage{dcolumn}
\usepackage{bm}
\usepackage{amssymb}
\usepackage{epstopdf}
\usepackage{color}
\usepackage{dsfont}
\usepackage{float}
\usepackage{physics}

\begin{document}

\title{On the possibility of a terahertz light emitting diode based on a dressed quantum well}

\author{S. Mandal$^{1}$}\email {subhaska001@e.ntu.edu.sg}
\author{K. Dini$^{1}$}
\author{O. V. Kibis$^{2,1}$}
\author{T. C. H. Liew$^1$}\email {tchliew@gmail.com}
\affiliation{$^1$Division of Physics and Applied Physics, School
of Physical and Mathematical Sciences, Nanyang Technological
University, Singapore 637371, Singapore}
\affiliation{${^2}$Department of Applied and Theoretical Physics,
Novosibirsk State Technical University, Karl Marx Avenue 20,
Novosibirsk 630073, Russia}


\begin{abstract}
We consider theoretically the realization of a tunable terahertz light emitting diode from a quantum well with dressed electrons placed in a highly doped p-n junction. In the considered system the strong resonant dressing field forms dynamic Stark gaps in the valence and conduction bands and the electric field inside the p-n junction makes the QW asymmetric.  It is shown that the electrons transiting through the light induced Stark gaps in the conduction band emit photons with energy directly proportional to the dressing field. This scheme is tunable, compact, and shows a fair efficiency.
\end{abstract}
\maketitle
\section{Introduction}
Terahertz radiation has widespread applications spanning the fields of (astro)physics, biology, medicine, and security \cite{Tonouchi_2007}. While semiconductors typically feature in the most compact optoelectronic systems, they are challenging to employ in the terahertz range at which electromagnetic transitions are usually absent. Exceptions have appeared in the use of transitions between exciton-polariton branches \cite{ISavenko_2011,Barachati_2015,Huppert_2014}, direct-indirect exciton transitions \cite{Kyriienko_2013,Kristinsson_2013}, and interband subband transitions \cite{Nathan_2014}, however, potential competition with larger quantum cascade lasers is undemonstrated. Cascade systems based on exciton-polaritons \cite{Liew_2013} and nanoparticles \cite{Arnardottir_2018}  were considered theoretically, but only operate at a fixed frequency. Ideally, a compact semiconductor system could be controlled by an external field to vary its operation frequency.

\begin{figure}[b]
\centering
\includegraphics[width=0.5\textwidth]{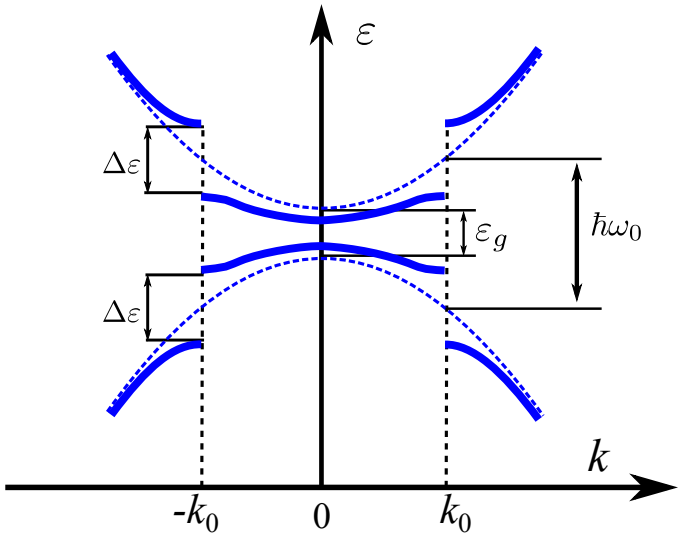}
\caption{Band structure  of the bare (dashed blue curves) and the dressed (solid blue curves) QW electronic states. Due to the mixing of bare valence and conduction bands, band gaps $\Delta\varepsilon$ open at the resonance point $k_0$ in $\mathbf{k}$ space.}
\label{fig:fig1}
\end{figure}
An intense enough electromagnetic field gives rise to a coupled light-matter object known as a dressed electron, which has been studied in various low dimensional systems, including: semiconductor quantum dots \cite{Koppenhofer_2016,Chestnov_2017}, quantum rings \cite{Kozin_2018,Sigurdsson_14}, quantum wells \cite{Wagner_10, Kibis_2012, Teich_13} etc. The appearance of band gaps due to the ac Stark effect in the band structure of various nanostructures  was first predicted theoretically \cite{Goreslavskii_1969} and later observed experimentally \cite{Vu_2004}.  The strong electromagnetic field mixes the valence and conduction bands of the system and consequently  band gaps $\Delta\varepsilon$, tunable in-situ by changing the intensity of the field, appear at the resonant points. In Figure~\ref{fig:fig1} a schematic diagram of the dynamic Stark gaps is shown. 

It was previously shown that  the dynamic Stark gaps in a quantum well with dressed electrons (dressed QW) can restrict the electron oscillation, which can be used  to realize a frequency comb when excited with sub-THz frequency \cite{Mandal_2018}. As the frequency of a dynamic Stark gap can be tuned into the THz range, it is natural to ask whether it can serve in the generation of THz radiation. This is far from obvious as in a normal QW electromagnetic transitions across a dynamic Stark gap are forbidden as there is no change in the symmetry between electrons in the states at the top and bottom of the gap \cite{Liberato_2013,Chestnov_2017}. However, as we will show, the use of an asymmetric QW alleviates this problem. Furthermore, by considering a geometry where the QW is placed between p and n doped semiconductors, we predict that it is possible to realize a pn junction \cite{Ando_1982,Nag_2000} operating at terahertz frequency. By considering the band-bending diagram and the transition matrix elements in the QW, we determine the relevant transition rates, and employ a simplified rate equation model to estimate the terahertz generation efficiency.
 
\section{Band bending of a quantum well in a pn junction}
Let us consider  an intrinsic QW sandwiched  between a p-n junction having doping concentrations $N_a$ and $N_d$ in the p and n sides, respectively. Following Ref.~\cite{Zeghbroeck_book4,Supp_mater} the potential variation for the electrons throughout the double heterojunction can be expressed as  
\begin{align}\label{potential_field}
&\phi(z)= 0,~\text{for}~z\le -(z_p+l/2).\nonumber\\
&\phi(z)= \frac{eN_a}{2\epsilon}\left[z^2+2z\left(z_p+\frac{l}{2}\right)+\left(z_p+\frac{l}{2}\right)^2\right],\nonumber\\
&~~~~~~~~~~~\text{for}~ -(z_p+l/2)\le z\le -l/2. \nonumber\\
&\phi(z)=\phi_p+\frac{e\left(N_h-N_e\right)}{2\epsilon} \left[-\frac{z^2}{l}+z+\frac{3l}{4}\right]\nonumber\\
&~~~~~~~~~~+\frac{eN_dz_n}{2\epsilon}(l+2z),~\text{for}~ -l/2\le z\le l/2. \nonumber\\
&\phi(z)=\phi_p+\phi_{QW}+\frac{eN_d}{2\epsilon}\left[z^2-2\left(z_n+\frac{l}{2}\right)z+z_nl+\frac{l^2}{4}\right],\nonumber\\
&~~~~~~~~~~~\text{for}~ l/2\le z\le z_n+l/2. \nonumber\\
&\phi(z)=\phi_p+\phi_{QW}+\phi_{n},~\text{for}~z\ge z_n+l/2.
\end{align}
\begin{figure}[b]
\includegraphics[width=0.5\textwidth]{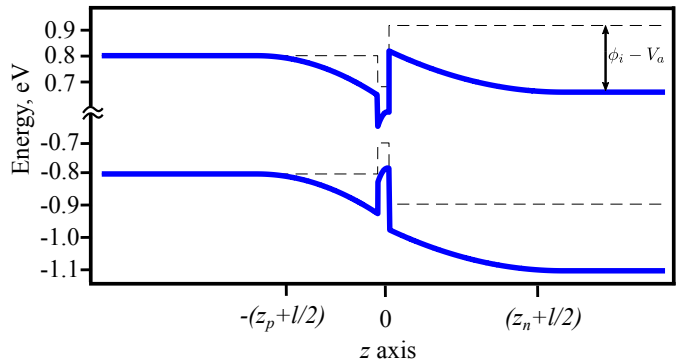}
\centering
\caption{Potential variation in the double heterojunction consisting of an Al$_{0.15}$Ga$_{0.85}$As p type layer, GaAs QW, and Al$_{0.2}$Ga$_{0.8}$As n type layer. The electric field inside the pn junction makes the QW asymmetric. The black dotted line indicates the potential of each layer before they are connected. Parameters: $N_a=2\times10^{16}$ cm$^{-3}$, $N_d=10^{16}$ cm$^{-3}$, $T = 4$ K, $l=8$ nm, $\varepsilon_g=1.4$ eV, $m_e=0.067m_0$, $m_h=0.47m_0$, $m_0$ being the free electron mass, $F_n=50$ meV, $F_p=2.5$ meV, $\phi_p=137$ V, $\phi_n=140$ V, $E^n_c=0.9$ eV, $E^p_v=-0.8$ eV.}
\label{fig:fig2}
\end{figure}
Here $e$ is the electronic charge; $\epsilon$ is the permittivity of the material; $z_p$ and $z_n$ are the widths of the depletion regions in the p and n side, respectively; $l$ is the thickness of the QW; $N_h$ and $N_e$ are the number of holes and electrons per unit area inside the QW; $\phi_p+\phi_{QW}+\phi_n=\phi_i-V_a$, is the potential difference between the p side and the n side; $\phi_i$ is the total built in potential; $\phi_p= eN_az_p^2/2\epsilon$, $\phi_{QW}=e(N_h-N_e)l/2\epsilon+eN_dz_nl/\epsilon$, and $\phi_n=eN_dz_n^2/2\epsilon$ are the built in potentials in the p side,  in the QW, and in the n side, respectively and $V_a$ is the externally applied voltage. The electron and hole density in the QW can be expressed as 
\begin{equation}
N_{e(h)}=\frac{m_{e(h)}}{\pi\hbar^2}k_BT\log\left[1+\exp\left(\frac{|F_{n(p)}|-|E_{c(v)}^0|}{k_BT}\right)\right].
\end{equation}
Here $m_e$ and $m_h$ are the effective electron and hole masses in the QW. $k_B$ is  Boltzmann's constant and $T$ is the temperature. $E_c^0$ and $E_v^0$ correspond to the minima and maxima of first electron and hole subbands of the QW, respectively.  $F_n$ and $F_p$ are the quasi Fermi levels in the conduction band and in the valence band of the QW, respectively, which can be expressed as \cite{joyce_dixon_1977,Zeghbroeck_book2}
\begin{align}
F_n= &E^n_c-e(\phi_p+\phi_{QW}+\phi_{n})+k_BT\Bigg{[}\ln\left(\frac{N_d}{N^n_c}\right)\nonumber\\
&+\frac{1}{\sqrt{8}}\left(\frac{N_d}{N^n_c}\right)-\left(\frac{3}{16}-\frac{\sqrt{3}}{9}\right)\left(\frac{N_d}{N^n_c}\right)^2+...\Bigg{]},\\
F_p=&-E^p_v-k_BT\Bigg{[}\ln\left(\frac{N_a}{N^p_v}\right)+\frac{1}{\sqrt{8}}\left(\frac{N_a}{N^p_v}\right)\nonumber\\
&-\left(\frac{3}{16}-\frac{\sqrt{3}}{9}\right)\left(\frac{N_a}{N^p_v}\right)^2+...\Bigg{]}.\label{QFL}
\end{align} 
Here $N^n_c=2({m_e} k_B T/2\pi\hbar^2)^{{3}/{2}}$ is the effective density of states at the conduction band edge of the n type semiconductor; $N^p_v=2({m_h} k_B T/2\pi\hbar^2)^{{3}/{2}}$ is the effective density of states at the valence band edge of the p type semiconductor; $E^n_c$ is the minimum of the conduction band in the n side; $E^p_v$ is the maximum of the valence band in the p side. Using Eqs.~(\ref{potential_field}-\ref{QFL}) the quasi Fermi levels of the QW can be controlled. The effect of $\phi(z)$ on the electron and hole potentials throughout the double heterojunction is plotted in Figure~\ref{fig:fig2} where we consider GaAs and AlGaAs as appropriate materials.
\section{QW energy spectrum under electromagnetic dressing}
We consider now that the system is subjected to a linearly polarized high frequency monochromatic electromagnetic field, $E\cos \omega_0 t$, which is linearly polarized along the $z$ direction (direction perpendicular to the QW plane) with the electric field amplitude $E$ and the frequency $\omega_0$. For simplicity, we have assumed that the electric field is spatially homogeneous. Assuming that the photon energy is greater than the light-matter interaction characteristic energy ($\hbar\omega_0\gg dE$, where d is the interband dipole moment), the corresponding Hamiltonian can be expressed as\cite{Kibis_2012,Janes_Cummings,Shammah_2015}
\begin{align}\label{Hamiltonian}
\hat{H}=&\hbar\omega_0\hat{a}^{\dagger}\hat{a}+\frac{\varepsilon^c(\mathbf{k})+\varepsilon^v(\mathbf{k})}{2}\hat{I}+\frac{\varepsilon^c(\mathbf{k})-\varepsilon^v(\mathbf{k})}{2}\hat{\sigma_z}\nonumber\\
&-id\sqrt{\frac{2\pi\hbar\omega_0}{V}}\left(\hat{\sigma}_+\hat{a}-\hat{\sigma}_-\hat{a}^{\dagger}\right).
\end{align}
The superscripts $c$ and $v$ correspond respectively to the conduction and valence bands; $\hat{a}$ and      $\hat{a}^{\dagger}$are photon creation and annihilation operators; and $\hat{\sigma}_\pm$  are the electron interband transition operators.

We choose the dressing field frequency $\omega_0$ such that $\hbar \omega_0>\varepsilon_g$, where  $\varepsilon_g$ is the band gap of the QW. Under this condition the dressing field strongly mixes the valence band and the conduction band of the quantum well and the stationary state of the system shows  energy gaps $\Delta \varepsilon$  at the resonant points of the Brillouin zone, $k_0$, where the energy of the photons $\hbar \omega_0$ matches with the energy difference between the bands (see Figure~\ref{fig:fig1}(a)) \cite{Goreslavskii_1969}. The modified energy spectrum $\varepsilon(\mathbf{k})$ that arises due to the mixing of the two bands $\varepsilon^c(\mathbf{k})$ and $\varepsilon^v(\mathbf{k})$  is expressed by \cite{Kibis_2012}
\begin{align}
\varepsilon(\mathbf{k})=&\frac{\varepsilon^c(\mathbf{k})+\varepsilon^v(\mathbf{k})}{2}\pm\frac{\hbar \omega_0}{2}\pm\frac{\alpha(\mathbf{k})}{2|\alpha(\mathbf{k})|}\sqrt{(\Delta\varepsilon)^2+\alpha^2(\mathbf{k})}. 
\end{align}
Here $\mathbf{k}=(k_x,k_y)$ is the in plane electron wave vector; $\varepsilon^c(\mathbf{k})=\hbar^2k^2/2m_e+\varepsilon_g/2$ is the first electron sub-band; $\varepsilon^v(\mathbf{k})=-\hbar^2k^2/2m_h-\varepsilon_g/2$ is the first hole sub-band;  $\alpha(\mathbf{k})=\varepsilon^c(\mathbf{k})-\varepsilon^v(\mathbf{k})-\hbar\omega_0$ is the resonance detuning; $\Delta\varepsilon=dE$ is the dynamic Stark gap. It should be noted that the dressing field amplitude is chosen in such a way that it satisfies the strong light-matter coupling condition and consequently is not absorbed. This condition can be expressed as $\Omega_R \tau_e\gg 1$, where $\Omega_R=\Delta\varepsilon/\hbar$ is the Rabi frequency of  interband electron  transitions and $\tau_e$ is the average lifetime of the electron states. 
\begin{figure}[b]
\centering
\includegraphics[width=0.5\textwidth]{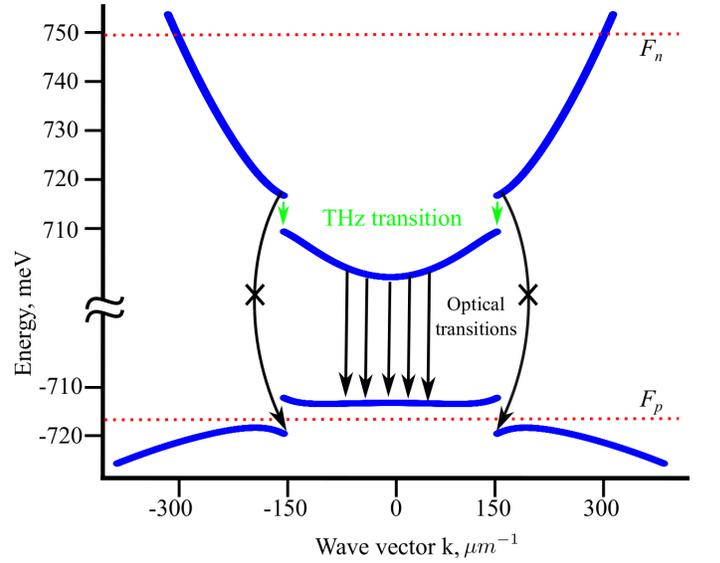}
\caption{Energy spectrum of the first electron and hole sub-band for a dressed GaAs QW. Since in GaAs based systems $m_h\gg m_e$ the upper dressed valence band appears flat. The two quasi Fermi levels, $F_n$ and $F_p$, are shown with red dashed lines. The green and black arrows indicate the THz and allowed optical transitions, respectively. The black curved arrows indicate the possible direct optical transition from the top conduction band to the bottom valence band which is prevented by setting $F_p$ at the Stark gap in the valence band. Parameters: $\hbar\omega_0=\varepsilon_g+15$ meV, $\Delta\varepsilon=8$ meV. All other parameters are kept the same as those in Figure~\ref{fig:fig2}.}
\label{fig:fig3}
\end{figure}
\section{THz transition matrix element and transition rate}
In order to estimate the THz rate we calculate the matrix element for the THz transition, defined as
\begin{equation}\label{ME1}
M_T= \bra{\psi_N^c(\mathbf{k})}\mathbf{r}\ket{\psi_N^c(\mathbf{k+\mathbf{q}})} .
\end{equation}
Here $\mathbf{q}$ is the wave vector of the emitted photon, $\ket{\psi_N^{c}(\mathbf{k})}$ and $\ket{\psi_N^{v}(\mathbf{k})}$ are the dressed conduction and valence band states with photon number $N$, respectively and expressed as \cite{Kibis_2012}
\begin{align}
\ket{\psi_N^{c}(\mathbf{k})}=&\sqrt{\frac{\Omega(\mathbf{k})+|\alpha(\mathbf{k})|}{2\Omega(\mathbf{k})}} \ket{\psi_c(\mathbf{k}),N}\nonumber\\
&+i\frac{\alpha(\mathbf{k})}{|\alpha(\mathbf{k})|}\sqrt{\frac{\Omega(\mathbf{k})-|\alpha(\mathbf{k})|}{2\Omega(\mathbf{k})}}\ket{\psi_v(\mathbf{k}),N+1}\\
\ket{\psi_N^{v}(\mathbf{k})}=&\sqrt{\frac{\Omega(\mathbf{k})+|\alpha(\mathbf{k})|}{2\Omega(\mathbf{k})}} \ket{\psi_v(\mathbf{k}),N}\nonumber\\
&+i\frac{\alpha(\mathbf{k})}{|\alpha(\mathbf{k})|}\sqrt{\frac{\Omega(\mathbf{k})-|\alpha(\mathbf{k})|}{2\Omega(\mathbf{k})}}\ket{\psi_c(\mathbf{k}),N-1}.
\end{align}
Here $\ket{\psi_{c}(\mathbf{k})}$ and $\ket{\psi_{v}(\mathbf{k})}$ are the bare conduction and valence band states, respectively; and $\Omega(\mathbf{k})=\sqrt{(\Delta\varepsilon)^2+\alpha^2(\mathbf{k})}$.
In a QW, the wave functions of the conduction and valence sub-bands can be written as
\begin{eqnarray}
\psi_c(\mathbf{k})=\psi_{1c}(z)u_c(\mathbf{r})e^{i\mathbf{k}\mathbf{r}}\\ 
\psi_v(\mathbf{k})=\psi_{1v}(z)u_v(\mathbf{r})e^{i\mathbf{k}\mathbf{r}}
\end{eqnarray}
where $u_{c}(\mathbf{r})$ and $u_{v}(\mathbf{r})$ are the Bloch functions of conduction and valence bands in a bulk semiconductor material; $\psi_{1c}(z)$ and $\psi_{1v}(z)$ are the envelope wave functions of the first conduction and valence sub-band arising from the quantization of transverse motion of electrons and holes in the QW.  We have to take into account that the Bloch functions, $u_{c,v}(\mathbf{r})$, oscillate with the crystal lattice period, $a$, whereas the characteristic scale of the wave functions $\psi_{1c,1v}(z)$ is the QW thickness, $l\gg a$. Taking the above mentioned conditions into account the interband dipole moment can be expressed as 
\begin{equation}\label{IBD}
d=\bra{\psi_{1c}(z)}ez\ket{\psi_{1v}(z)}.
\end{equation}
Since the dipole operator $(e\mathbf{r})$ will only act on the electronic part of the wave function and the photon states corresponding to different photon number are orthogonal, the  THz matrix element in Eq.~(\ref{ME1}) can be approximated to 
\begin{align}\label{ME2}
M_{T}\simeq\frac{1}{2}\left[\bra{\psi_{1c}(z)}z\ket{\psi_{1c}(z)}-\bra{\psi_{1v}(z)}z\ket{\psi_{1v}(z)}\right].
\end{align}
This corresponds to the emitted photons having linear polarization along $z$ axis. The matrix elements corresponding to photons having polarization in the other two directions (i.e. along $x$ and $y$) vanish. For a symmetric QW both the terms in Eq.~(\ref{ME2}) will go to zero separately, but for an asymmetric QW the wave functions $\psi_{1c}(z)$ and $\psi_{1v}(z)$ do not have a definite parity resulting in nonzero $M_T$. Using our parameters we have found that $M_T\simeq 1~$nm which is similar to the one obtained in Ref.~\cite{Nathan_2014}. The THz transition rate can be expressed as 
\begin{align}\label{THz_rate}
W_{T}(\omega_{T} )  &= \frac{4e^2|M_{T}|^2\sqrt{\epsilon_r}}{\hbar \pi \epsilon_0 c^3 } \rho_{T}(\omega_{T} ) \omega^3_{T}.
\end{align}
Here, ${\epsilon_r}$ is the dielectric constant of the QW, $\epsilon_0$ is the free space permittivity, $c$ is the velocity of light in free space, and $\rho_{T}(\omega_{T} )$ is the total number of electronic states in the upper conduction band participating in the THz transition with frequency  $\omega_{T}$. In order to realize a monolithic THz source we consider placing the system inside a  THz cavity. It is possible to use metal microcavities where the light is confined at sub-wavelength scales in the THz frequency range and the dispersion of the THz photons can be flat \cite{Todorov_2010}. This also allows us to neglect THz transitions other than the one that is resonant with the cavity. The presence of the cavity modifies the three dimensional continuum of photonic modes to a two  dimensional continuum of photonic modes and Eq.~(\ref{THz_rate}) becomes \cite{Kakazu_1994}
\begin{align}\label{Cavity_THz_rate}
W_{T}(\omega_{T} )  &= \frac{4e^2|M_{T}|^2}{\hbar \pi \epsilon_0 c^2L_c } \rho_{T}(\omega_{T} ) \omega^2_{T}.
\end{align}
Here $L_c$ is the length of the cavity. Due to the presence of $\mathbf{q}$, we choose to calculate $\rho_{T}$ numerically using,
\begin{align}
\rho_{T}(\omega_{T} )=2\sum_{\mathbf{k_i}}\delta(\varepsilon_i(\mathbf{k_i})-\varepsilon_f(\mathbf{k_i}+\mathbf{q})-\hbar\omega_{T})
\end{align}
where the summation runs over all the initial states $\mathbf{k_i}$. Each electron making the THz transition can also undergo an optical transition directly to the lowest valence band as shown with the  dashed arrows in Figure~\ref{fig:fig3}. This undesired transition can be stopped by setting $F_p$ at the Stark gap (see Figure~\ref{fig:fig3}). It should be noted that the transition from the top conduction band to the top valence band is restricted by momentum conservation. In order to estimate the efficiency of the device we solve the rate equations for different energy bands coupled with the THz photons inside the cavity,
\begin{align}\label{Rate_Eqn}
\frac{\partial n_u}{\partial t}&=R_{in}-W_{T}n_u\left(n_{t}+1\right)+W_{T}n_ln_{t}-\frac{n_u}{\tau_e}\nonumber\\
\frac{\partial n_l}{\partial t}&=W_{T}n_u\left(n_{t}+1\right)-W_{T}n_ln_{t}-\frac{B}{V}n_lp-\frac{n_l}{\tau_e}\nonumber\\
\frac{\partial n_t}{\partial t}&=W_{T}n_u\left(n_{t}+1\right)-W_{T}n_ln_{t}-\frac{n_t}{\tau_t}
\end{align}
\begin{figure}[t]
\centering
\includegraphics[width=0.5\textwidth]{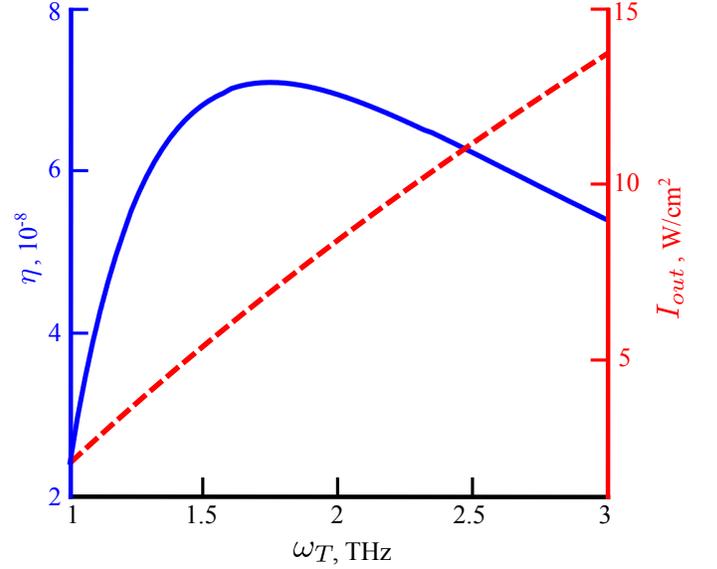}
\caption{Variation of the device efficiency, $\eta$ (blue line), and output intensity, $I_{out}$ (dashed red line),  as a function of the emitted THz photon frequency from a dressed QW with sides 2.5 mm. Parameters: $L_c=0.8~\mu$m, $\tau_e=0.1$ ns, $\tau_t=0.5$ ps, $B=2\times10^{-10}$ cm$^3$/s.}
\label{fig:fig4}
\end{figure}
\begin{figure*}[t]
\centering
\includegraphics[width=0.9\textwidth]{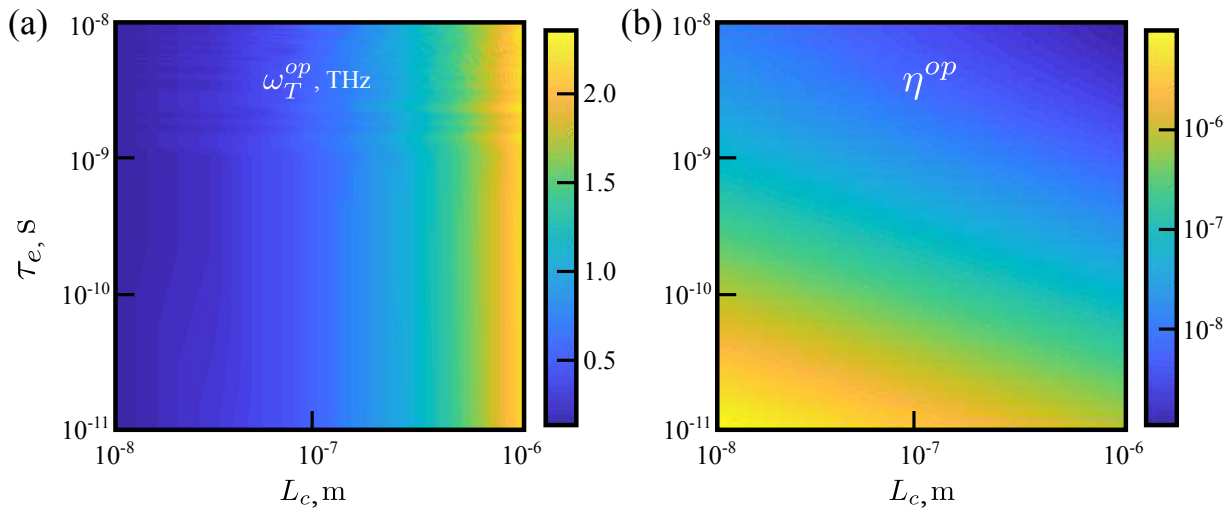}
\caption{Plot of the optimum THz photon frequency in (a) and its corresponding efficiency in (b), as  functions of cavity length, $L_c$, and average electronic lifetime, $\tau_e$. The values of $L_c$ are taken from Ref.\cite{Todorov_2010}. All the parameters are kept the same as those in Figure~\ref{fig:fig4}.}
\label{fig:fig5}
\end{figure*}

where $n_u$ and $n_l$ are the number of the electrons in the upper and lower conduction band of the QW, respectively;  $n_t$ and $\tau_t$ are the number of THz photons inside the cavity and their lifetime, respectively; $B$ is the radiative recombination rate in usual QWs, which has the typical value $2\times10^{-10}$ cm$^3$/s \cite{Ridley_1990}; $V$ is the volume of the QW; $p$ is the number of holes in the QW, which is calculated from $F_p$; $\tau_e$ is the average lifetime of the electrons, which we take as $0.1$ ns; $R_{in}$ is the rate at which the electrons are injected into the QW. In principle, the THz transitions corresponding to $\hbar\omega_T>\Delta\varepsilon$ can be considered, which requires the probability of occupation for each state in the conduction band, however this is beyond the scope of the present work. We limit ourselves to the THz transitions between the states that lie near the Stark gap such that $\hbar\omega_T\simeq\Delta\varepsilon$. This allows us to assume that, due to the fast relaxation of the electrons, the lower energy states in the upper conduction band are filled and the higher energy states in the lower conduction band are empty.  Electron-electron scattering is likely to increase  with the increase in temperature, which will put the system out of the strong coupling regime. To discard any temperature effect we consider that the system is in a low temperature environment such that $k_BT\ll \Delta\varepsilon$, which can be  achieved using liquid $^4$He. The input intensity ($I_{in}$) is defined as
\begin{align}
I_{in}&=\frac{eR_{in}V_a}{A}+\frac{1}{2}\varepsilon_0cE^2\label{I_in}.
\end{align}
Here $V_a={(F_n-F_p)}/{e}$, and  the first and second term in the right-hand side of Eq.~(\ref{I_in}) represent the intensity due to the electric current in the device and the dressing field intensity, respectively. The output intensity ($I_{out}$) is defined as
\begin{align}
I_{out}&=\frac{1}{A}\left({n_t}/{\tau_t}\right)\hbar\omega_T\label{I_out},
\end{align}
where, the term ${n_t}/{\tau_t}$ represents the rate of emission of the THz photons from the cavity. In a GaAs based QW the interband dipole moment, $d \simeq10$ D, and  a Stark gap of meV order corresponds to $E\simeq 10^5$ V/cm \cite{Goreslavskii_1969}. The efficiency of the device, defined as, $\eta=I_{out}/I_{in}$, is calculated by finding the steady state of  the rate equations in Eq.~(\ref{Rate_Eqn}), which is obtained in a self consistent way where for each THz frequency the total number of electrons in the QW (defined by the $F_n$) is kept constant by varying $R_{in}$. In Figure~\ref{fig:fig4} the efficiency and output intensity are plotted as a function of THz frequency. The efficiency shows a maximum near 1.5 THz. This corresponds to the Pauli exclusion principle. Once all the electronic states in the lower conduction band are filled the population of the THz photons does not increase even after increasing $\omega_T$. Although $I_{out}$ increases linearly as a function of $\omega_T$, $I_{in}$ increases quadratically resulting in a maximum in the efficiency. It should be noted that $\omega_T$ can be easily tuned by changing the intensity of the dressing field. As shown in Figure~\ref{fig:fig5}, the optimum frequency, $\omega_T^{op}$, of the THz emission can be controlled mainly by varying $L_c$ with $\tau_e$ having very little effect on it. However, the optimum efficiency, $\eta^{op}$, decreases with the increase of $L_c$ and $\tau_e$. This is understandable, as from Eq.~\ref{Cavity_THz_rate} it is clear that THz transition rate, $W_T$, decreases with the increase of $L_c$. On the other hand, the average electronic lifetime, $\tau_e$, vary depending upon all the possible non-radiative scattering processes. Smaller $\tau_e$ creates vacancy in the lower conduction band increasing the THz transition probability, which results in higher $\eta^{op}$.  The typical efficiency of the device, $\eta\simeq 7\times 10^{-6}$, which is comparable to other state-of-the-art proposals for THz emission from Stark split bands \cite{Nathan_2014}. The main energy cost in the considered system derives from the second term in Eq.~(\ref{I_in}) (right-hand side), that is, the energy needed to supply the dressing field. As this energy is not absorbed by the system, one could consider that this component of the input energy can be recycled in principle. In such an idealistic case, the efficiency would become $\eta\simeq 5\times 10^{-3}$. One can also define the quantum efficiency, $\zeta=\left({n_t}/{\tau_t}\right)/R_{in}$, which represents the number of THz photons emitted per injected electron. The typical quantum efficiency of the device, $\zeta\simeq0.5$, which is very close to unity.
\section{Conclusion}
To conclude, we have considered theoretically a THz source based on  a dressed QW  inside a highly doped p-n junction coupled to a THz cavity. By solving the rate equations we showed that the system subjected to a forward bias voltage can act as a  THz emitting diode having energy efficiency of $7\times 10^{-6}$ and quantum efficiency of 0.5. The advantages of our scheme is that it is composed of only one quantum well, making it more compact and easier to fabricate, and the device is tuneable such that it can give different THz frequencies.
\section{Acknowledgments}
S.M., K.D. and T.L. were supported by the Ministry of Education (Singapore), grant 2017-T2-1-001. O.K. was supported by the Russian Foundation for Basic Research (project 17-02-00053) and Ministry of Science and High Education of Russian Federation (projects 3.4573.2017/6.7 and 3.8051.2017/8.9)


\end{document}